# Amplitude Modulation Effects in Cardiac Signals

Randall Peters[1], Erskine James[2] & Michael Russell[3]

[1]Physics Department and [2]Medical School, Department of Internal Medicine
Mercer University, Macon, GA
[3]Physiology Department
MCG/UGA Medical Partnership
Athens, GA.

**Abstract**

A subject's heart beat can be nearly invisible in a spectrum, when that spectrum is generated using conventional methods of Fourier analysis. The phenomenon has been observed in records of both electrocardiography type and seismocardiography type. The mechanisms of nonlinear physics responsible for these complexities involve the phenomenon of amplitude modulation. Fortunately, there is a simple remedy to prevent loss of valuable frequency domain information. Instead of operating on the raw signal, one simply rectifies that signal before performing the fast Fourier transform (FFT) calculation. Alternatively, nearly equivalent spectra can be obtained by operating on the signal with the Teager-Kaiser operator before doing the FFT.

**Background**

An electrocardiograph (ECG) recording depicts the electrical activity of the human heart and provides information on timing defects and ischemic injury [1]. Lead I from a typical three-lead ECG represents left-to-right current flow across the heart. In addition to the usual time domain graph, various graphs of frequency domain type, generated from the time record, are expected to become a valuable supplementary tool with which to identify heart abnormalities.

Though valuable information concerning the heart may be available from frequency domain records obtained from Lead I, it is shown later in this article that significant parts of the ECG signal can be nearly invisible when conventional Fourier analysis is used, due to the nonlinear phenomenon of amplitude modulation. Fortunately, there is a simple remedy for this loss of frequency domain information. Removal of certain wave parts, before doing the FFT, appears to restore the missing periodicity information. Specifically, culling by rectification the parts that correspond to current flow away from the positive electrode (negative to the isoelectric line), restores information concerned with the primary ventricular depolarization. Though the origin of the masking process is not understood in detail, it is postulated that it may result from the diffuse atrial repolarization that occurs simultaneous with ventricular depolarization. Further analysis of multiple leads may be required to isolate the source of the amplitude modulation.

Seismocardiography is the study of accelerations produced by movement of heart muscle and blood through the large arteries[2]. Although left-ventricular contraction (identified by the ECG R-wave) generally is cause for significant chest-wall acceleration at the frequency of the heart-beat, it is also accompanied by multiple-mode vibrations that typically peak around 10 Hz. As with the waves other than R-type in the ECG, these additional seismocardiographic modes tend to make the heartbeat and respiration invisible in an FFT based directly on the signal. As with the ECG, the otherwise missing low-frequency spectral lines can be recovered by means of demodulation using either rectification or the T-K operator.

**Nature of Amplitude Modulation**

Fig. 1 shows why the `information' component(s) of an amplitude modulated carrier (upper left) is not immediately obvious in the Fourier transform of the signal (upper right). It is present in the form of sidebands close to the spectral line of the carrier, as illustrated in Fig. 2. Normally, however, a `close-up' of any spectral line is not displayed, rather only a broad-band log-abscissa plot like the upper right graph of Fig. 1. Note that in these figures, and others that follow, graph ordinate-units are either relative-linear or relative-logarithmic.

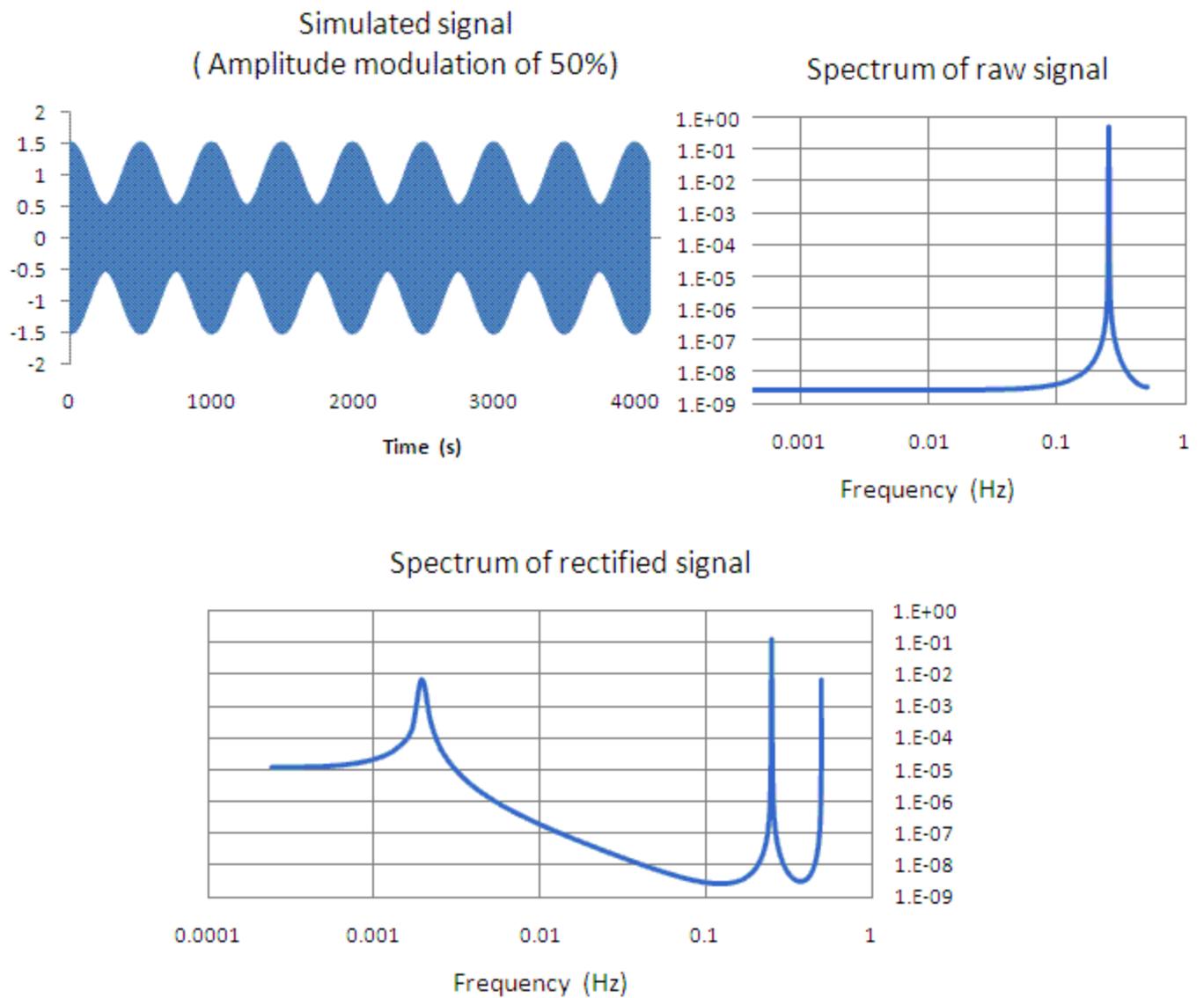

**Figure 1.** Simulation of a near-monochromatic carrier (frequency $f_c = 0.25$ Hz) that is amplitude modulated by a near-monochromatic `information' component (frequency $f_m = 0.002$ Hz).

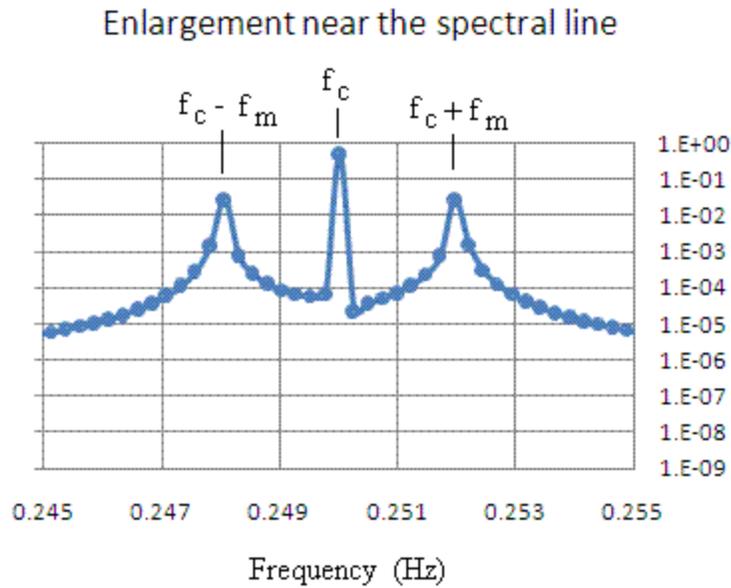

**Figure 2.** Close-up of the upper-right plot of Fig. 1, showing that the information is present in the form of sidebands.

Figures 1 and 2 are readily understood from theory by considering the equation that is the math-basis of the figures; i.e.,

$$y(t) = (A + M \cos \omega_m t) \sin \omega_c t \qquad (1)$$

where $\omega_c = 2\pi f_c$, $\omega_m = 2\pi f_m$, $A = 1$, and $M = 0.5$. By means of a trigonometric identity, Eq.(1) can be shown to be equivalent to

$$y(t) = A \sin \omega_c t + 0.5 M [\sin(\omega_c + \omega_m)t + \sin(\omega_c - \omega_m)t] \qquad (2)$$

which shows why there is no readily perceived (`stand-alone') spectral line corresponding to $f_m$.

**Remedy**

Rather than work with multiple spectra, with at least one being narrow-band to address the sidebands (as in Fig. 2), it is more convenient to take advantage of the principle of AM-radio that was perfected more than a century ago. `Extraction' of $f_m$ is accomplished by working with the rectified signal. Here, rectified means `half-wave', where the negative-going parts of the signal are `clipped-off' (reset to zero). In full-wave rectification, the absolute value of the signal would be instead used.

It is seen from the lower graph of Fig. 1 that rectification results in a `stand-alone' spectral line of frequency $f_m$. It has the disadvantage, however, of generating the spurious spectral line (0.5 Hz in the figure) whose frequency is twice that of $f_c$. This is inconsequential in radio work, since the frequencies that are significantly higher than $f_m$ are removed by low-pass filtering.

**Example Cardiac Records**

**Electrocardiogram case**

The first example to illustrate this amplitude modulation effect in some heart-related records is one involving an electrocardiogram, with results shown in Fig. 3

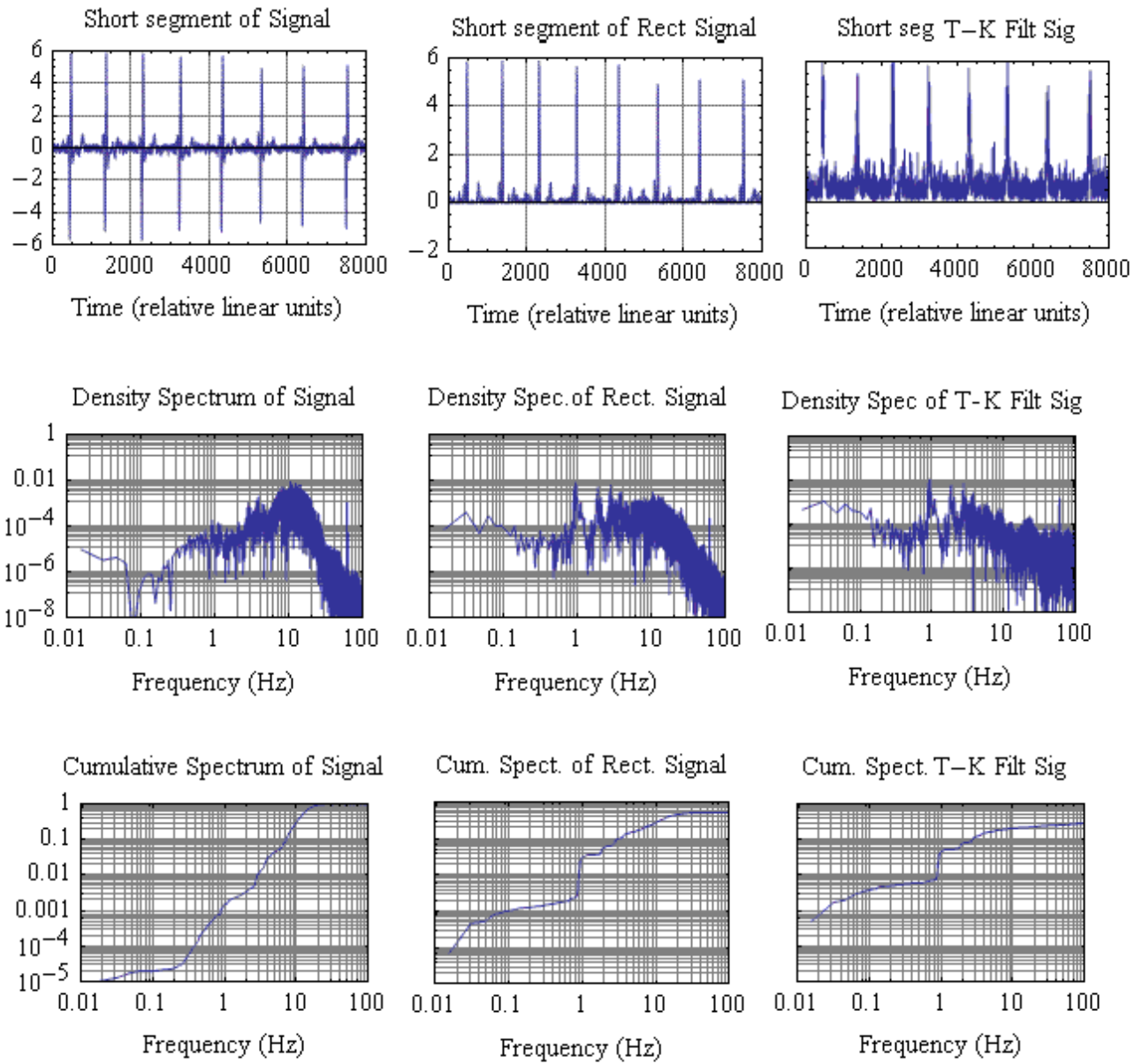

**Figure 3.** Graphs of an ECG. A short section of the temporal record is shown in the top row of graphs. The density spectrum for each of the first row cases is provided in the 2nd triplet of graphs. Finally, cumulative spectra for each case are provided in the 3rd row.

Heartbeat periodicity is unmistakeable in the time plots, of which the raw signal shows clearly the well-known P, Q, R, S, and T waves. The cumulative spectrum is much less `cluttered' than the spectral density, and so it is especially useful for studies of this type [3]. A spectral `line' in a density plot corresponds to a `step' in the associated (ever increasing) cumulative plot.

The following is especially worthy of note. In spite of the obvious periodic features in the time plots of Fig. 3, the heartbeat rate of approximately 1 Hz (60 beats/min), does not show up in the spectra generated from the raw signal (left column, rows 2 and 3). It is clearly revealed, however, in the spectra generated from the rectified signal (middle column, rows 2 and 3). The same is also true for the cases (right column) in which the signal was treated with the Teager-Kaiser operator before doing the FFT [4]. This means that the P, Q, R, S, and T events that define the signal are acting collectively to form an amplitude modulated signal.

It should be noted that the T-K operator being presently used is slightly different than what has been historically used. The discrete form of historical type is one using

$$E_n = x_n^2 - x_{n-1} \cdot x_{n+1} \qquad (3)$$

Because this expression is quadratic in form, the T-K operator is frequently referred to as an energy operator. The continuous case from which this discrete form is derived is equal to the peak value of twice the sum of kinetic and potential energies (i.e., total energy) per unit mass, only if x corresponds to position. For the experiments described in the present article, x corresponds to the derivative of acceleration; which in engineering terminology is called `jerk'. To make the time plots appear more similar in shape to those of the half-wave rectified signal, the square root of $E_n$ has been plotted in all of the figures. Whereas the continuous form of the T-K operator can never be negative (since the energy is positive definite), the discrete form can be less than zero for cases of slow sample rate looking at rapidly changing x. For the FFT calculations, the Mathematica code used in this work accounts automatically for any imaginary components that result when the argument of the square root is negative. In the case of the plots, any complex values are ignored and only the real values are graphed.

**Finger Plethysmograph case**

It is interesting to compare this ECG case with similar calculations performed on data obtained with a finger plethysmograph, shown in Fig. 4.

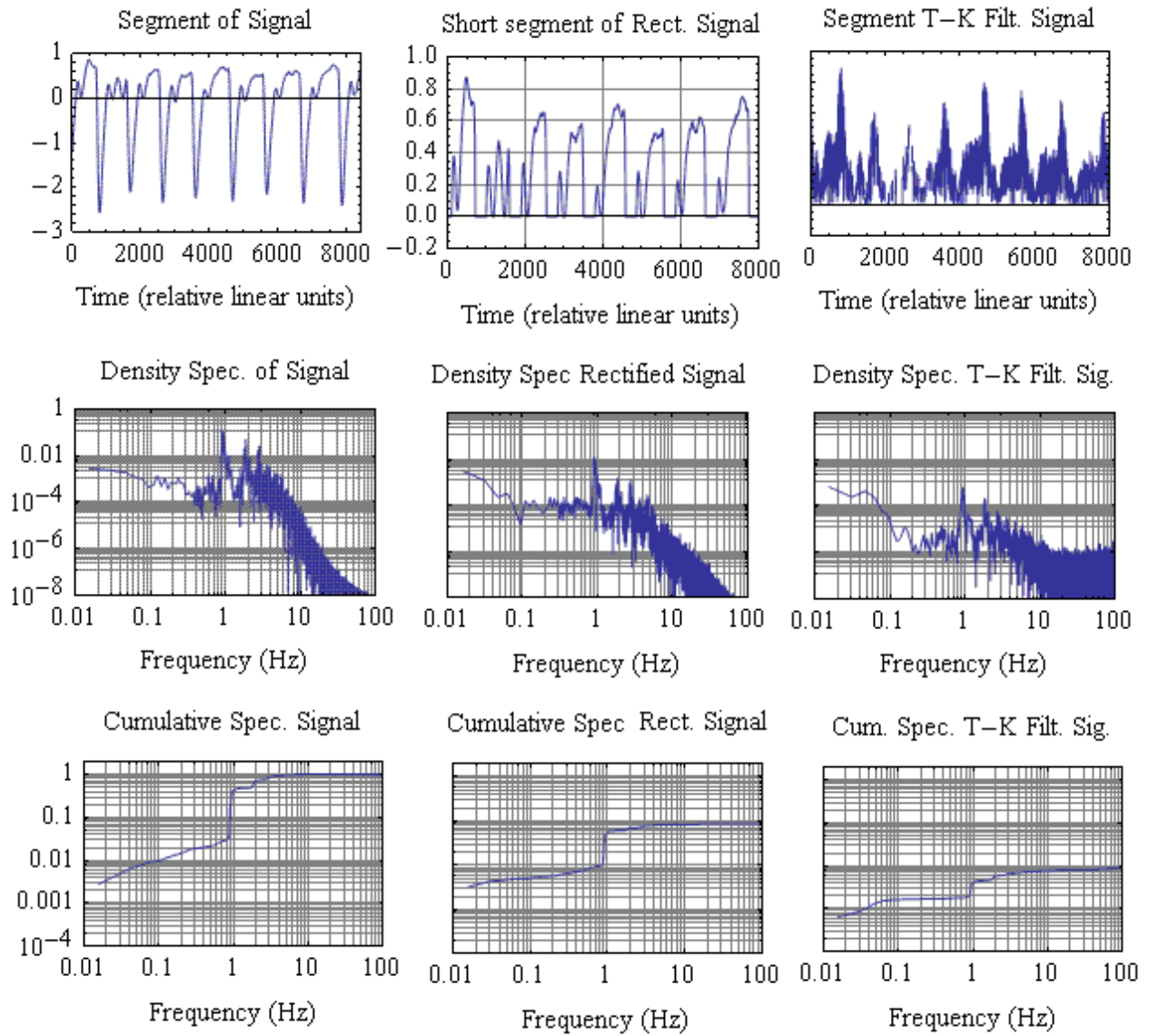

**Figure 4.** These graphs involving finger plethysmograph data show no features of amplitude modulation. The T-K filter output is seen to be very noisy when the raw signal contains rapidly changing components.

**Seismocardiography case**

For all the seismocardiograph cases presented in this article, geophone data was acquired, using Windaq software, at an ADC sample rate of 98 per second. Thus the Nyquist frequency was 49 Hz, which is seen to be consistent with the maximum frequency shown in the frequency domain plots. In the ECG and Plethysmograph cases of the earlier figures, completely different (commercial) instrument packages were used and the sample rate there was 1000 per s. In the graphs for those cases, the maximum frequency shown is 100 Hz, which is well below the

Nyquist frequency of 500 Hz. Nothing of significance is to be realized by including the additional range from 100 to 500.

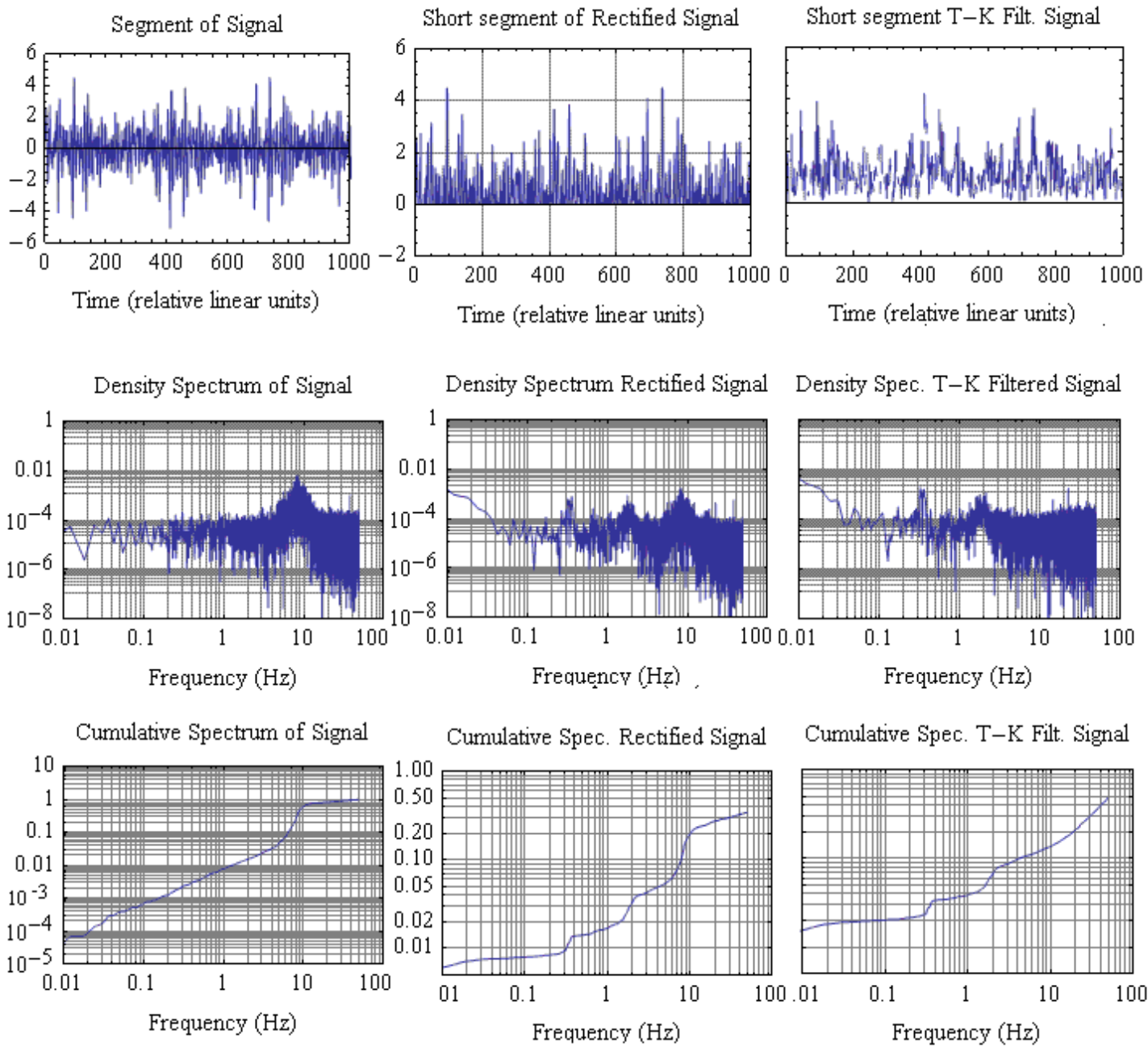

**Figure 5.** Seismocardiograph records frequently show amplitude modulation features, as deduced from the plots of this figure. Whereas the FFT of the raw signal does not reveal them, both heartbeat periodicity (at approx. 110 beats/min) and respiration periodicity (at approx. 20 inspirations/min) become visible in the spectra generated from the demodulated raw signal.

## Advantage of the Cumulative Spectrum

Because the cumulative spectrum is much less cluttered than the density spectrum from which it is obtained by integration, multiple plots can be usefully overlaid as shown in Fig. 6.

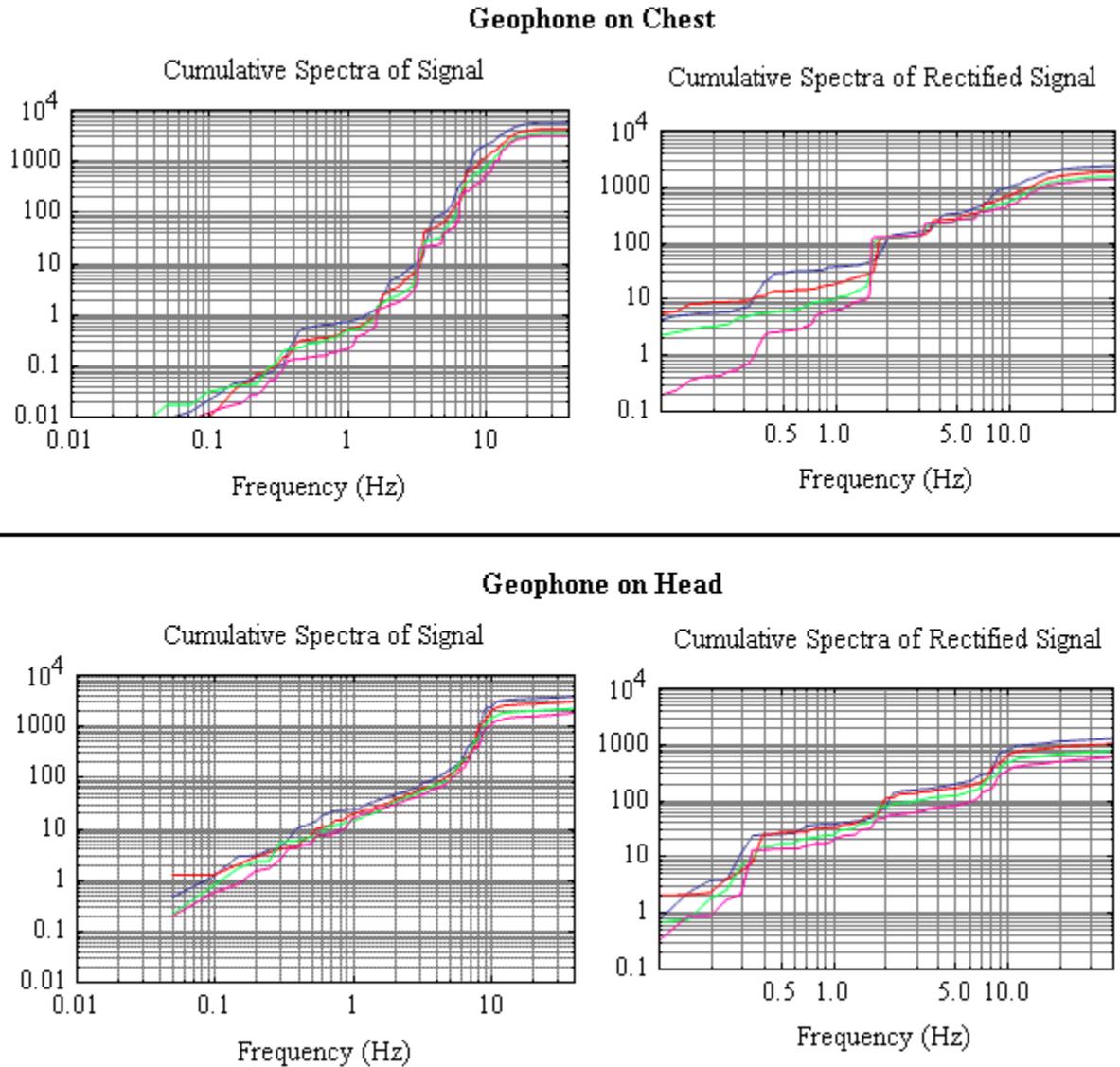

**Figure 6.** Example of overlaying multiple plots of the cumulative spectrum in a single graph. The four plots for a given case correspond to four sequential, contiguous time segments, whose ordering are blue, red, green, and magenta.

In each of the two cases of Fig. 6, the data was recorded with a geophone accelerometer, the output of which is proportional to velocity, because of the Faraday-law detector employed. The data collection in each case followed immediately after moderate exercise by the subject. Thus from the time of the first plot (blue) to that of the last plot (magenta), a slowing of the heart-rate can be recognized. The total time of each of the records was of the order of two to three minutes.

The human eye/brain has difficulty with the quantitative comparison of two different graphs, except when they are overlaid as in Fig. 6. Such figures are expected to facilitate the recognition of heart abnormalities involving variable rates.

**Conclusion**

It is clear from the data of this paper that two different time records should be generally considered, when it comes to the calculation of cardiac spectra. The possibility may exist for information loss, due to amplitude modulation effects, when only the raw signal is treated. Such loss can be avoided by looking also at the rectified signal.

In generating the 2nd time record, either rectification or the T-K operator can be used to condition the raw signal before doing the FFT. For most cases less information is available from the latter compared to the former, as is especially evident by comparing their cumulative spectra shown in Fig. 5.

**Acknowledgment**

This work was funded by the Heart, Lung, and Blood Institute of the National Institutes of Health- project title ``Three-dimensional cardiac accelerometry for cardiac monitoring'', grant number 1R15HL 103489-01.